\documentclass[%
 reprint,
superscriptaddress,
amsmath,
amssymb,
aps,
pra,
floatfix
]{revtex4-1}
\usepackage{cancel}
\usepackage{titlesec}
\usepackage{braket}
\usepackage{amsmath}
\usepackage{amssymb}
\usepackage{graphicx}
\usepackage{dcolumn}
\usepackage{bm}

  \newcommand{\e}{\epsilon}

\titlespacing*{\section}
{0pt}{5.5mm}{4mm}
\titlespacing*{\subsection}
{0pt}{3mm}{3mm}
\begin{document}


\title{Influence of Vacuum modes on Photodetection}

\author{S. A. Wadood}
\email{swadood@ur.rochester.edu}
 \affiliation{The Institute of Optics, University of Rochester,Rochester, NY 14627, USA.}
\affiliation{Center for Coherence and Quantum Optics, University of Rochester,Rochester, NY 14627, USA.}
\author{J. T. Schultz}
 \affiliation{The Institute of Optics, University of Rochester,Rochester, NY 14627, USA.}
\affiliation{Center for Coherence and Quantum Optics, University of Rochester,Rochester, NY 14627, USA.}
\author{A. Nick Vamivakas}
 \affiliation{The Institute of Optics, University of Rochester,Rochester, NY 14627, USA.}
\affiliation{Center for Coherence and Quantum Optics, University of Rochester,Rochester, NY 14627, USA.}
\affiliation{Department of Physics and Astronomy, University of Rochester, Rochester, NY 14627, USA.}
\author{C. R. Stroud, Jr.}
 \affiliation{The Institute of Optics, University of Rochester,Rochester, NY 14627, USA.}
\affiliation{Center for Coherence and Quantum Optics, University of Rochester,Rochester, NY 14627, USA.}
\affiliation{Department of Physics and Astronomy, University of Rochester, Rochester, NY 14627, USA.}



\date{\today}

\begin{abstract}
Photodetection is a process in which an incident field induces a polarization current in the detector. The interaction of the field with this induced current excites an electron in the detector from a localized bound state to a state in which the electron freely propagates and can be classically amplified and detected. The induced current can interact not only with the applied field, but also with all of the initially unpopulated vacuum modes. This interaction with the vacuum modes is assumed to be small and is neglected in conventional photodetection theory. We show that this interaction contributes to the quantum efficiency of the detector. We also show that in the Purcell enhancement regime, shot noise in the photocurrent depends on the bandwidth of the the vacuum modes interacting with the detector. Our theory allows design of sensitive detectors to probe the properties of the vacuum modes.
\end{abstract}

\pacs{Valid PACS appear here}
\maketitle


\section{\label{sec:level1}Introduction}
Conventional photodetection theory as formulated by Glauber \cite{Glauber}, Mandel, and others \cite{Kelley} has been remarkably successful at describing a wide variety of phenomena observed in quantum optics such as the Hanbury Brown--Twiss effect \cite{HBT}, bunching and antibunching \cite{MandelAntibunching}, Hong--Ou--Mandel interferometry \cite{HongOuMandelInterference}, etc. According to this theory the fluctuating vacuum cannot lead to any changes in the expected photocurrent. This is to be expected as there are no states of lower energy than the vacuum into which the field could decay \cite{milonni_photodetection_and_causality_vacuum_contribution}. However, expectations of the photocurrent are not the only properties of the system that can be measured. One can also measure the noise in the observed signal as well as various correlations. We explore the question: does the presence of initially unexcited vacuum modes of the field play a role in these quantities?
\par These vacuum modes have been used to explain a number of observed effects in quantum optics including spontaneous emission \cite{Purcell}, the Lamb shift \cite{Lamb_Shift,weltonLambshift}, the Casimir effect \cite{Casimir}, induced coherence in down-conversion \cite{MandelInducedCoherence,MenzelInducedCoherence}, and, more recently, noise in electro-optic sampling of THz pulses \cite{THz_vacuum}. It has been suggested \cite{bs_mirror}, but never directly observed that ``vacuum ports'' introduced by beam splitters can increase the noise in homodyne detector experiments.
\par In a conventional detector, the back-action of the detection process on the field is neglected. The field induces a dipole current in the detector, which is damped by the excitation of an electron into a continuum. There cannot be spontaneous absorption of photons from the vacuum and hence no vacuum contribution to the excitation rate. Thus, normal ordering appears as the natural formalism to explain photodetection as photon absorption or as ``clicks'' on a detector.\par
However, the induced current \textit{can} act as a source term in Maxwell's equations and thus interact with the vacuum modes at the detector. This is expected to be a small, but finite, effect for a good detector, which has small back-action on the field. In this paper, we present a detector model which includes the interaction of the dipole current with the vacuum modes.\par
In Sec. II, the equations of motion and the approximations used to obtain them are explained. In Sec. III, the equations of motion are used to find the mean current and show how the vacuum modes can affect the quantum efficiency of the detector. Sec. 4 deals with current fluctuations, their relation to the vacuum reservoir bandwidth, and issues in detector design to enhance the effect of vacuum modes. Sec. 5 summarizes and concludes the paper.
\begin{figure}[h!]
\includegraphics[width=\columnwidth]{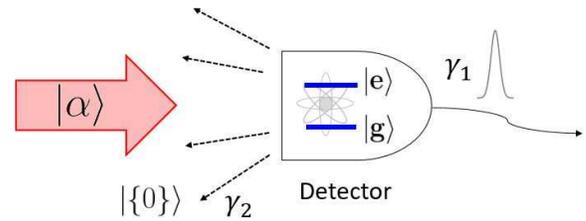}
\caption{Model for photodetection. A coherent state $\ket{\alpha}$ illuminates a two-level detector in the ground state. The two-level system is coupled to two reservoirs, an electronic reservoir and a radiative reservoir of vacuum modes denoted by $\ket{\{0\}}$. $\gamma_{1(2)}$ is the damping constant associated with the electronic (vacuum) reservoir. The figure shows a classical pulse being generated in the electronic reservoir and dashed lines showing photons scattering into the vacuum modes.}
\label{fig:detectormodel}
\end{figure}
\section{Hamiltonian and equations of motion}
We model our detector as a two-level system coupled to field and electronic reservoirs at zero temperature. The total Hamiltonian is given by the  sum of the detector and reservoir Hamiltonians and the interaction Hamiltonians:
\begin{align} 
H&=H_{0}+H_{int}\\
H_{0}&=\hbar w_{e}b^{\dagger}_{\e}b_{\e}+\sum_{k}\hbar w_{k}a^{\dagger}_{k}a_{k}+\sum_{l}\hbar w_{l}c^{\dagger}_{l}c_{l}\\
H_{int}&=-\sum_{k}\hbar g_{k}(a_{k}b^{\dagger}_{\e}+b_{\e}a^{\dagger}_{k}+a_{k}b_{\e}+b^{\dagger}_{\e}a^{\dagger}_{k})\nonumber\\
&-\sum_{l}\hbar g_{l}(b^{\dagger}_{\e}c_{l}+c^{\dagger}_{l}b_{\e}+b_{\e} c_{k}+c^{\dagger}_{k}b^{\dagger}_{\e}),
\end{align}
where $b_{\e}=\ket{g}\bra{\epsilon}$ is the transition-projection operator from the ground state $\ket{g}$ to excited state $\ket{\epsilon}$ with frequency $w_\epsilon$, and $a_{k}$($c_{k}$) is the annihilation operator for the $k^{th}$ field (electronic) mode with coupling frequency $g_{k}$. The ground state energy of the detector is taken to be zero. The detector and all the reservoir oscillators are assumed to be in the ground state at $t=0$. The detector is illuminated with a laser represented by a coherent state $\ket{\alpha}$. The initial state of the system is then $\ket{0}_{det}\otimes\ket{\alpha}\otimes\ket{\{0\}}_{vac}\otimes\ket{\{0\}}_{elec}$, where $det,vac,elec$ refer to detector state and vacuum and electronic reservoirs respectively and $\otimes$ denotes the tensor product.\par
Since $[b_{\e}(t),a_{k}(t)]=[b_{\e}(t),c_{k}(t)]=0$ for all k, we can write the equal time products of detector and reservoir operators in arbitrary order. We have chosen normal ordering for the present calculation. The Heisenberg equation of motion for $b_{\e}$, under the rotating wave Hamiltonian is
\begin{align}
\dot{b}_{\e}&=-i w_{\e}b_{\e}+i\sum_{k}g_{k,\e}([b_{\e},b^{\dagger}_{\e}]a_{k})+i\sum_{l}g_{l,\e}([b_{\e},b^{\dagger}_{\e}]c_{l})\nonumber\\
&=-i w_{\e}b_{\e}+i\sum_{k}g_{k,\e}(\ket{g}\bra{g}-\ket{\e}\bra{\e})a_{k}\nonumber\\
&\hspace*{+4mm}+i\sum_{l}g_{l,\e}(\ket{g}\bra{g}-\ket{\e}\bra{\e})c_{l},\nonumber\\
&=-i w_{\e}b_{\e}-i\sum_{k}g_{k,\e}\sigma_{z,\e}a_{k}-i\sum_{l}g_{l,\e}\sigma_{z,\e}c_{l},
\end{align}
where $\sigma_{z,\e}=\ket{\e}\bra{\e}-\ket{g}\bra{g}$. The equations for $a_{k}$ and $c_{l}$ are similar such that
\begin{align} 
\dot{a}_{k}&=-iw_{k}a_{k}+ig_{k,\e}b_{\e},\\
\dot{c}_{l}&=-iw_{l}c_{l}+ig_{l,\e}b_{\e},
\end{align}
\par Substituting the formally integrated equation for $a_{k}$ and $c_{l}$ into the equation for $b_{\e}$ gives
\small
\begin{align} 
&\dot{b}_{\e}=-i w_{\e}b_{\e}-i\sum_{k}g_{k,\e}\sigma_{z,\e}a_{k}(0)e^{-iw_{k}t}-i\sum_{l}g_{l,\e}\sigma_{z,\e}c_{l}(0)e^{-iw_{l}t}\nonumber\\
&\hspace*{+7mm}+\sum_{k}g_{k,\e}g_{k,\e}\sigma_{z,\e}\int_{0}^{t}dt'b_{\e}(t')e^{-iw_{k}(t-t')}\nonumber\\
&\hspace*{+7mm}+\sum_{l}g_{l,\e}g_{l,\e}\sigma_{z,\e}\int_{0}^{t}dt'b_{\e}(t')e^{-iw_{l}(t-t')}.
\end{align}
\normalsize
We can use the equality $\sigma_{z,\e}b_{\e}=-b_{\e}$. In photodetection, one is interested in the weak excitation limit where there is negligible population in the excited state since the excited electrons are pulled into the detector circuit by a bias voltage. We can therefore approximate the commutator as
\begin{align}
[b_{\e},b^{\dagger}_{\e}]=\ket{g}\bra{g}-\ket{\e}\bra{\e}\approx\ket{g}\bra{g}.
\end{align}
Similarly the closure relation for the detector Hilbert space can also be approximated as
\begin{align}
\hat{I}=\ket{g}\bra{g}+\ket{\e}\bra{\e}\approx\ket{g}\bra{g}\approx[b_{\e},b^{\dagger}_{\e}],
\end{align}
which shows that in the weak excitation regime the detector behaves like a harmonic oscillator \cite{Mollow}.\\
If we make a Markov approximation \cite{milonnibook,MandelandWolf} for both the reservoirs we have the equation
\begin{align}\label{eqn::forb_e}
\dot{b}_{\e}&=-(iw_{\e}+\gamma_{\e})b_{\e}+i\sum_{k}g_{k,\e}a_{k}(0)e^{-iw_{k}t},
\end{align}
where $\gamma_{\e}=\gamma_{1}+\gamma_{2}$ and $\gamma_{1(2)}=\frac{\pi}{2}g^2_{1(2)}(\e)\rho_{1(2)}(\e)$ is the detector decay rate into the electronic (field) reservoir and $\rho_{1(2)}(\e)$ is the density of states of the corresponding reservoir around frequency $\e$. In the Markov approximation, the two independent reservoirs contribute to two distinct decay rates. In particular, the damping constant $\gamma_{2}$ appearing in eq. (\ref{eqn::forb_e}) represents the interaction of the detector energy level with the vacuum modes, and can be modified by effects such as Purcell enhancement that will be considered later. We note that there is an equivalent interpretation of this interaction in terms of radiation reaction. It was shown in \cite{milonni1,senitzky1973radiationandvacuum,milonnibook} that spontaneous emission linewidth and the Lamb shift in the projection operator and the atomic inversion operator can be attributed either to source-field back-action or to the vacuum field, depending on whether we use normal or antinormal ordering in the interaction Hamiltonian.\par
The solution for $b_{\e}$ and $a_{k}$ is then given by
\small
\begin{align}\label{eqn:1_transitionoperatorsolutions}
b_{\e}(t)&=b_{\e}(0)e^{-(iw_{\e}+\gamma_{\e})t}+i\sum_{k}g_{k\e}a_{k}(0)f_{k}(t)+i\sum_{l}g_{l\e}c_{l}(0)f_{l}(t),
\end{align}
\begin{align}\label{eqn:2_transitionoperatorsolutions}
a_{k}(t)&=a_{k}(0)e^{-iw_{k}t}+b_{\e}(0)h_{k\e}(t)-\sum_{k'}p_{kk'\e}a_{k'}(0),
\end{align}
\normalsize
where
\begin{align}
f_{k}(t)&=\int_{0}^{t}dt'e^{-iw_{k}t'-(iw_{\e}+\gamma_{\e})(t-t')},\\
h_{k\e}(t)&=ig_{k\e}\int_{0}^{t}dt'e^{-(iw_{\e}+\gamma_{\e})t'-iw_{k}(t-t')},\\
p_{kk'\e}&=g_{k\e}g_{k'\e}\int_{0}^{t}dt'f_{k'\e}(t')e^{-iw_{k}(t-t')},
\end{align}
and the solution for $c_{k}(t)$ is same as that of $a_{k}(t)$. The function $f_{k}$ affects the Langevin noise interacting with the detector, whereas the function $p_{kk'\e}$ represents the effect of the source field mode $k'$ generated by the level $\ket{\e}$ on the field or electronic mode $k$. Note that we treat the electronic reservoir as bosonic even though the photoelectrons are fermions. The use of a fermionic reservoir does not alter our results in a fundamental way, as the langevin equations for $c_{k}$, the mean current and fluctuations remain the same \cite{FermionicLangevinEquationgardiner}. \par 
We can use the solutions of $b_{\e}$, $a_{k}$ and $c_{k}$ to calculate observable quantities such as mean and noise of the photocurrent.
\section{Mean Current}
We are interested in the rate of excitation of quanta in the electronic reservoir i.e. $\braket{\hat{i}}=\sum_{l}\partial_{t}\braket{c^{\dagger}_{l}c_{l}}=\sum_{l}\Braket{\dot{n}_{l}}$. For a coherent state incident on the detector, the mean current in the electronic reservoir is then given by:
\begin{align}{\label{eqn::meancurrent}}
\braket{\hat{i}}&=\sum_{l}\Braket{\dot{n}_{l}}=\sum_{l}ig_{l}(\braket{c^{\dagger}_{l}b_{\e}}-\braket{b^{\dagger}_{\e}c_{l}})\nonumber\\
&=2|\alpha|^2\text{Re}\left\{\sum_{l}g_{l}g_{L}f_{L}p^{*}_{lL}\right\}\nonumber\\
&=2|\alpha|^2\text{Re}\left\{g^{2}_{L}f_{L}\int_{0}^{t}f^{*}_{L}(t')(\sum_{l}g^{2}_{l}e^{iw_{l}(t-t')}dt')\right\}\nonumber\\
&=2|\alpha|^2\gamma_{1}g^{2}_{L}|f_{L}(t)|^2
\end{align}
where $L$ indicates the laser mode, the detector and both reservoirs are assumed to be in the ground state at $t=0$, and we have used $\sum_{l}g^{2}_{l}e^{iw_{l}(t-t')}=2\gamma_{1}\delta(t-t')$ in the last step. In the steady state limit, all the transients die out and the mean current is given as 
\begin{align}
\braket{\hat{i}}=\frac{2\gamma_{1}}{(\gamma_{1}+\gamma_{2})^2+(w_{L}-w_{\e})^2}g_{L}^2|\alpha|^2.
\end{align}
The mean current in the electronic reservoir depends on the intensity of the laser, the strength of the dipole moment between the detector's ground and excited state, and on the two decay rates of the excited states. If the laser is resonant with the detector, the factor of $\frac{\gamma_{1}}{(\gamma_{1}+\gamma_{2})^2}$ gives the branching ratio of the two reservoirs. This effect can be interpreted as a contribution to the detector quantum efficiency. A detector with shorter excited state radiative lifetimes will tend to scatter photons, yielding a lower quantum efficiency. However, the magnitude of the effect would depend on the ratio $\xi=\frac{\gamma_{2}}{\gamma_{1}}$. At resonance we have
\begin{align}{\label{eqn::meancurrent2}}
\braket{\hat{i}}=\frac{1}{(1+\xi)^2}\frac{2g_{L}^2|\alpha|^2}{\gamma_{1}}=\frac{1}{(1+\xi)^2}\braket{\hat{i}}_{\text{n.o}},
\end{align}
where $\Braket{i}_{\text{n.o}}$ is the current expected from normally ordered statistics. In general there will be other loss mechanisms $\xi_{i}=\frac{\gamma_{i}}{\gamma_{1}},i \neq 1$ for an $i^{th}$ reservoir in the detector and then $\xi=\sum_{i}\xi_{i}$. A high quantum efficiency detector requires a smaller $\xi$, such that all the excited electrons are captured by the electronic reservoir. For a low quantum efficieny such that $0.1\leq\xi\leq1$, $\gamma_{2}$ can be modified using the Purcell effect, provided other loss mechanisms are weaker and do not contribute to $\xi$.
\par
The mean current is affected by the vacuum modes through the quantum efficiency. How is the current noise affected by the vacuum? The next section addresses this question.
\section{Current Fluctuations}
The square of the electronic current operator is given as
\begin{align}
\hat{i}^2&=
\sum_{\scriptscriptstyle{l,l'}}
g_{l}g_{l'}(b^{\dagger}_{\e}c_{l}c^{\dagger}_{l'}b_{\e}+a^{\dagger}_{l}b_{\e}b^{\dagger}_{\e}c_{l'}),
\end{align}
where we have used $b^{\dagger}_{\e}b^{\dagger}_{\e}=b_{\e}b_{\e}=0$.
Using the commutators $c_{l}c^{\dagger}_{l'}=c^{\dagger}_{l'}c_{l}+\delta_{ll'}$ and $b_{\e}b^{\dagger}_{\e}\approx 1+b^{\dagger}_{\e'}b_{\e}$, we have
\begin{align}
\hat{i}^2&=\sum_{\scriptscriptstyle{l,l'}}g_{l}g_{l'}(2b^{\dagger}_{\e}c^{\dagger}_{l'}c_{l}b_{\e}+b^{\dagger}_{\e}b_{\e}\delta_{ll'}+c^{\dagger}_{l}c_{l'}),
\end{align}
where we have swapped the dummy index $l$ with $l'$ to add the first two terms. The expectation value is
\begin{align}
\braket{\hat{i}^2}&=\sum_{l,l'}2|\alpha|^4g^2_{L}g_{l}g_{l'}f^{*}_{L}f_{L}p^{*}_{lL}p_{l'L}\nonumber
\\&+\sum_{l,l'}|\alpha|^2g_{l}g_{l'}p^{*}_{lL}p_{l'L}+\sum_{l}g^{2}_{l}|\alpha|^2g^2_{L}|f_{L}|^2
\end{align}
Similarly, we find $\braket{\hat{i}}^2$ to be
\begin{align}
\braket{\hat{i}}^2&=|\alpha|^4g^2_{L}\bigg(\sum_{l}(f^{*}_{L}p_{lL}+f_{L}p^{*}_{lL})\bigg)^2\nonumber\\
&=\sum_{l,l'}2|\alpha|^4g^2_{L}g_{l}g_{l'}f^{*}_{L}f_{L}p^{*}_{lL}p_{l'L}
\nonumber\\&+\sum_{l,l'}2|\alpha|^4g^2_{L}g_{l}g_{l'}Re\{f^{2}_{L}p^{*}_{lL}p_{l'L}\}.
\end{align}
The variance is given as
\begin{align}
(\Delta \hat{i})^2&=\braket{\hat{i}^2}-\braket{\hat{i}}^2\\
&=\sum_{l,l'}|\alpha|^2g_{l}g_{l'}p^{*}_{lL}p_{l'L}+\sum_{l}|\alpha|^2g^{2}_{l}g^2_{L}|f_{L}|^2\nonumber\\&-\sum_{l,l'}2|\alpha|^4g^2_{L}g_{l}g_{l'}Re\{f^{2}_{L}p^{*}_{lL}p_{l'L}\}.
\end{align}
 In the Markov approximation, $\sum_{l}g^2_{l}=\rho g^{2}_{\e}\int dl=2\gamma_{1}\int \frac{dl}{\pi}$. The variance is
\begin{align}
(\Delta \hat{i})^2&=(\frac{\Omega}{\pi}+\frac{\gamma_{1}}{2})\braket{\hat{i}}-\frac{\braket{\hat{i}}^2}{2}\\
&\approx\frac{\Omega}{\pi}\braket{\hat{i}}, \text{for $\Omega\gg \gamma_{1}$} 
\end{align}
where $\Omega$ is the bandwidth of the electronic reservoir which is assumed to be much larger than $\gamma_{1}$ in the Markov approximation. As would be expected for shot noise, the current noise depends directly on the interaction bandwidth of the detector and electronic reservoir. The term $\braket{i}^2/2$ is a correction of $O(g^4_{L})$ and can be neglected. The only vacuum contribution to the noise is through the quantum efficiency factor which appears in the mean current.\par
Measurement of temporal coherence and squeezing requires knowledge of the two time current correlation, which is related to second order coherence properties of the field. Like the variance, the two time correlation function is only affected by the branching ratio of the vacuum and electronic reservoir, i.e. the quantum efficiency. The details of this calculation are given in the appendix. Moreover, the current mean and variance vanish if the laser is turned off, i.e., $\alpha=0$, and no energy is absorbed from the vacuum.\par
The extension of the detector from a two-level-system to a continuum of excited state levels is straightforward. Since we assume that the detector is never saturated, each level in the continuum is independent of the other and the cross-talk can be neglected. Then one can sum over $\e$ in the equations, with the coupling constant $g_{k}$ replaced by $g_{k\e}$. The fundamental results regarding the quantum efficiency still hold.
\section{Experimental Challenges}
We now discuss the challenges in realizing a detector that shows modification of photocurrent by changing remote boundary conditions. In the ideal and somewhat simpler case, such a detector would show a measurable difference of quantum efficiency in a cavity versus free space. The magnitude of the effect of modifying the vacuum modes would depend on the parameter $\xi$, introduced in eq. (\ref{eqn::meancurrent2}). In the `bad cavity' limit, and ignoring all other non-radiative losses, we have
\begin{align}\label{eqn::branchingratioparameter}
\xi=\frac{\gamma_{2}}{\gamma_{1}}=\frac{g^2_{k\e}}{\kappa \gamma_{1}}
\end{align}
where $g_{k\e}$ is the coupling of the vacuum mode of frequency $k=\frac{w_{\e}}{c}$ to the detector and $\kappa$ is the cavity linewidth or the bandwidth of the vacuum mode reservoir \cite{CavityPurcellEnhancement1987,harocheexploringthequantum}. In writing eq. (\ref{eqn::branchingratioparameter}), we have assumed that emission rate into the cavity mode is much larger than emission rate into modes not supported by the cavity. For $\xi\ll1$, which corresponds to an efficient detector, no change in quantum efficiency will occur according to eq. (\ref{eqn::meancurrent2}). Therefore, a `bad' detector is more likely to show a quantum efficiency change in the cavity. For semiconductors, $\gamma_{1}$ is inversely related to the transit time of the electrons, or the slower holes, from the point of excitation to the electrode that finally registers a click. This transit time can be controlled via the bias voltage and spatial properties of the detector to be in the range of 1ps to 1$\mu$s \cite{SalehFundamentalsofPhotonics}. On the other hand, $\gamma_{2}$ can be engineered by changing the bandgap and/or type of material used. However, any other non-radiative recombination mechanism like phonon scattering, Auger processes, or defect capture will add to the losses, lowering the value of $\xi$ and precluding the effect of the change in vacuum modes \cite{SalehFundamentalsofPhotonics}.\par
Eq. (\ref{eqn::branchingratioparameter}) suggests that for detectors with $\xi\approx1$, if $\kappa$ is changed adiabatically, the shot noise level in sensitive homodyning \cite{Shapiro,raymerhomodyneultrafastpulses} and squeezing experiments\cite{bs_mirror,CarmichaelNormallyOrderedShotNoise} can be modified.\par
\section{Conclusion}
Photodetectors act as probes for the electromagnetic field. The detector's induced dipole interacts not only with the illuminating field, but also with the vacuum modes. We have shown that this interaction affects the quantum efficiency of the detector. Furthermore, in the bad cavity limit, the shot noise and correlation of the photocurrent depends on the bandwidth of the vacuum mode reservoir. Even though our results are entirely based on a quantized field treatment, a classical analogy nevertheless exists; the modification of vacuum reservoir is analogous to changing the mutual impedance of an antenna in free space \cite{AntennaPurcell}.\par
Conventionally, vacuum modes have been probed through changes in excited state lifetimes of emitters coupled to the vacuum reservoir \cite{kleppnerpurcell,PurcellSuperconducting,Purcell,harocheexploringthequantum}. However, our results suggest that vacuum modes can affect dynamics of absorption processes like photodetection. Modifying the coupling constant $g_{k\e}$ of an excited state level $\e$ with the $k^{th}$ vacuum mode will affect both the mean and noise of the photocurrent. This allows for the design of photodetectors and cavity geometries that can be sensitive probes of changes in the quantum vacuum.
\section{Acknowledgements}
We would like to thank Peter Milonni and Gary Wicks for insightful discussions and the Army Research Office for the funding support under grant no. W911NF1610162.
\section{Appendix}
The two-time correlation function of the current for $\tau>0$ can be found using the solutions given by eqns. (\ref{eqn:1_transitionoperatorsolutions}-\ref{eqn:2_transitionoperatorsolutions}) and, after a lengthy but straightforward calculation, is found for $t_{2}=t_{1}+\tau$ as
\begin{widetext}
\begin{align}
&\Braket{\hat{i}(t_{1})\hat{i}(t_{2})}=\sum_{k,k'}g_{k}g_{k'}\bigg[\bigg(\sum_{k_{1}-k_{4}}g_{k_{2}}g_{k_{3}}\tilde{p}^{*}_{kk_{1}}(t_{1})\tilde{f}^{*}_{k_{2}}(t_{2})\tilde{f}_{k_{3}}(t)\tilde{p}_{k'k_{4}}(t_{2})
\Gamma^{(2,2)}_{k_{1}k_{2}k_{3}k_{4}}(t_{1},t_{2})+C.C\bigg)\nonumber\\
&+\bigg(\sum_{k_{1}-k_{4}}g_{k_{3}}g_{k_{4}}\tilde{p}^{*}_{kk_{1}}(t_{1})\tilde{p}^{*}_{k'k_{2}}(t_{2})\tilde{f}_{k_{3}}(t_{1})\tilde{x}_{k_{4}\e}(t_{1},t_{2})\Gamma^{(2,2)}_{k_{1}k_{2}k_{3}k_{4}}(t_{1},t_{2})+C.C\bigg)\nonumber\\
&+\sum_{k_{1}k_{2}}\tilde{p}^{*}_{kk_{1}}(t_{1})\tilde{p}_{k'k_{2}}(t_{2})\Gamma^{(1,1)}_{k_{1},k_{2}}(t_{1},t_{2})[b_{\e}(t_{1}),b_{\e}^{\dagger}(t_{2})]+\sum_{k_{1}k_{2}}g_{k_{1}}g_{k_{2}}\tilde{f}^{*}_{k_{1}}(t_{1})\tilde{f}_{k_{2}}(t_{2})\Gamma^{(1,1)}_{k_{1},k_{2}}(t_{1},t_{2})[c_{k}(t_{1}),c^{\dagger}_{k'}(t_{2})]\nonumber\\
&+i\sum_{k_{1}k_{2}}g_{k_{2}}\tilde{p}^{*}_{kk_{1}}(t_{1})\tilde{f}_{k_{2}}(t_{2})\Gamma^{(1,1)}_{k_{1},k_{2}}(t_{1},t_{2})[b_{\e}(t_{1}),c^{\dagger}_{k'}(t_{2})]
-i\sum_{k_{1}k_{2}}g_{k_{1}}\tilde{f}^{*}_{k_{1}}(t_{1})\tilde{p}_{k'k_{2}}(t_{2})\Gamma^{(1,1)}_{k_{1},k_{2}}(t_{1},t_{2})[c_{k}(t_{1}),b^{\dagger}_{\e}(t_{2})]
\bigg],
\end{align}
\end{widetext}
where \small
$x_{k\e}(t_{1},t_{2})=\int_{0}^{t_{2}-t_{1}}e^{-iw_{k}(t_{1}+t')-(iw_{\e}+\gamma_{\e})(t_{2}-t_{1}-t')}dt'$, $\tilde{p}_{k_{1}k_{2}}(t)=p_{k_{1}k_{2}}(t)e^{iw_{k_{2}}t}$, $\tilde{f}_{k}(t)=f_{k}(t)e^{iw_{k}t}$, $\Gamma^{(2,2)}_{k_{1}k_{2}k_{3}k_{4}}(t_{1},t_{2})=\braket{a^{\dagger^{(0)}}_{k_{1}}(t_{1})a^{\dagger^{(0)}}_{k_{2}}(t_{2})a^{(0)}_{k_{3}}(t_{2})a^{(0)}_{k_{4}}(t_{1})}$, $\Gamma^{(1,1)}_{k_{1},k_{2}}(t_{1},t_{2})=\braket{a^{\dagger^{(0)}}_{k_{1}}(t_{1})a^{(0)}_{k_{2}}(t_{2})}$\normalsize are the time ordered, normally ordered \cite{Glauber} second and first order coherence functions respectively for $t_{2}>t_{1}$ and $a^{(0)}_{k}(t)=a_{k}(0)e^{-iw_{k}t}$ are free-field annihilation operators. Note that all the commutators in the expression are complex functions and have no operator characteristics \cite{MandelandWolf}.\par
In the long-time limit in which the transients die out and the correlation function is stationary, we have for $\tau>0$
\begin{align}\label{eqn::currentcorrelation}
\Braket{\hat{i}(t)\hat{i}(t+\tau)}&\approx g^2_{L}|\alpha|^2\frac{e^{-iw_{L}\tau}}{(\gamma_{\e})^{2}+w_{L\e}^2}(\gamma^2_{1}e^{iw_{\e}\tau-\gamma_{\e}\tau}\nonumber\\
&\hspace*{+4mm}+\chi_{e}(\tau))+O(g^4_{L}),
\end{align}
where $w_{L\e}=w_{L}-w_{\e}$ and $\chi_{e}(\tau)$ is the electronic reservoir correlation function and behaves like a delta function in the Markov approximation. The last term of $O(g^4_{L})$ in eq. (\ref{eqn::currentcorrelation}) contains the second order field coherence functions.


%

\end{document}